\documentclass{aa}  

\usepackage{graphicx}
\usepackage{txfonts}
\usepackage{amsmath}    
\usepackage{amssymb}    
\usepackage{natbib,twoopt}
\bibpunct{(}{)}{;}{a}{}{,}             
\usepackage[breaklinks=true]{hyperref}
\hypersetup{
    colorlinks=true,
    linkcolor=blue,
    filecolor=magenta,      
    urlcolor=cyan,
    citecolor=blue
   }
\begin{document}

   \title{An empirical nonlinear power spectrum overdensity response}

    \author{G\'abor R\'acz\inst{1,2},
            Istv\'an Szapudi\inst{3}
            \and
            Istv\'an Csabai\inst{1}
          }
    \institute{Department of Physics of Complex Systems, ELTE E\"{o}tv\"{o}s Lor\'and University, 
                Pf. 32, H-1518 Budapest, Hungary\\
                \email{gabor.racz@ttk.elte.hu}
        \and
                Jet Propulsion Laboratory, California Institute of Technology, 4800 Oak Grove Drive, Pasadena, CA, 91109, USA\\
         \and
             Institute for Astronomy, University of Hawaii, 2680 Woodlawn Drive, Honolulu, HI, 96822, USA\\
             }

    \date{Received December 19, 2021}
    

  \abstract
  {The overdensity inside a cosmological sub-volume and the tidal fields from its surroundings affect the matter distribution of the region. The resulting difference between the local and global power spectra is characterized by the response function. }
   { Our aim is to provide a new, simple, and accurate formula for the power spectrum overdensity response at highly nonlinear scales based on the results of cosmological simulations and paying special attention to the lognormal nature of the density field.}
   {We measured the dark matter power spectrum amplitude as a function of the overdensity ($\delta_W$) in $N$-body simulation subsamples. We show that the response follows a power-law form in terms of $(1+\delta_W)$, and we provide a new fit in terms of 
   the variance, $\sigma(L)$, of a sub-volume of size $L$.}
   {Our fit has a similar accuracy and a comparable complexity to second-order standard perturbation theory on large scales, but it is also valid for nonlinear (smaller) scales, where perturbation theory needs higher-order terms for a comparable precision.
   Furthermore, we show that the lognormal nature of the overdensity distribution causes a previously unidentified bias: the power spectrum amplitude for a subsample with an average density is typically underestimated by about $-2\sigma^2$. Although this bias falls to the sub-percent level above characteristic scales of $200\textnormal{Mpc}h^{-1}$, taking it into account improves the accuracy of estimating power spectra from zoom-in simulations and smaller high-resolution surveys embedded in larger low-resolution volumes.}
   {}


   \maketitle
%

\section{Introduction}

   Two-point statistics of the cosmic density field are the principal tools for constraining cosmological models. Large-scale galaxy surveys, such as the Sloan Digital Sky Survey \citep{2004ApJ...606..702T}, 2MASS \citep{2001astro.ph..9403A}, APM Galaxy Survey-2 \citep{1994MNRAS.267..323B}, 2dF \citep{2001MNRAS.327.1297P}, BOSS \citep{2013AJ....145...10D}, DESI \citep{2016arXiv161100036D}, Pan-STARRS \citep{2016arXiv161205560C}, or the DES \citep{2005astro.ph.10346T}, provide data for such measurements. Even larger surveys are planned for the near future, for example Euclid \citep{2020A&A...643A..70T}, WFIRST/Roman \citep{2012arXiv1208.4012G}, SPHEREx \citep{2014arXiv1412.4872D}, LSST/Rubin \citep{2009arXiv0912.0201L}, and the Subaru Prime-Focus Spectrograph \citep{2016SPIE.9908E..1MT}. While some day we might map the cosmological density field of the observable Universe, at present these wide field surveys are complemented with narrower deep surveys, such as COSMOS \citep{2007ApJS..172....1S} and H20 \citep{2020MNRAS.493.2318B}, the deep survey mode of the HSC \citep{2018PASJ...70S...4A}. 
The geometry and volume of the survey window and the super-survey modes modulate the interpretation of any measured two-point statistics. 
The effects of the survey window are described in detail in \cite{1995clun.conf...13V} and \cite{2013IJAA....3..243S}. The effects of super-survey modes are usually treated as overdensity and tidal effects. These effects are discussed extensively in the literature \citep{2013PhRvD..87l3504T,2014PhRvD..89h3519L,2018PhRvD..97f3527A, 2018PhRvD..97d3532C, 2018JCAP...06..015B, 2019A&A...624A..61L, 2019MNRAS.490.4688R, 2019JCAP...10..004D, 2020JCAP...10..007C} based on standard perturbation theory (SPT) or the halo model, both of which use the matter bispectrum in the squeezed limit to calculate the overdensity response of the power spectrum \citep{2015JCAP...08..042W}. While the effect is significant even in large volumes \citep{2018JCAP...10..053B, 2019MNRAS.482.4253T}, it becomes particularly large when only small volumes are available -- as a rule of thumb, when the linear size of the survey window is less than a few hundred $\textnormal{Mpc}h^{-1}$.

The low-order responses in SPT are accurate on large scales, where the density field is only mildly non-Gaussian (where $k < 0.3 \textnormal{Mpc}h^{-1}
$ at $z=0$). Even second-order perturbation theory (PT) is limited in accuracy for smaller scales. In particular, the overdensity field itself has a non-Gaussian, approximately lognormal distribution, as shown by \cite{1991MNRAS.248....1C}.

Beyond SPT and the halo model, separate universe simulations have also been used to calculate the response functions numerically \citep{2015MNRAS.448L..11W,  2015JCAP...08..042W, 2019MNRAS.488.2079B}. These are useful for calculating the nonlinear overdensity responses and approximate tidal responses; for the latter this is done by adding anisotropic expansion or external (low-order) tidal fields \citep{2020MNRAS.496..483M, 2021MNRAS.503.1473S, 2021JCAP...04..041A}. Complex tidal fields and the density distribution are not taken into account in these simulations, since they are not embedded in a larger cosmological volume.

The state-of-the-art methods mentioned above provide precise results in most cases at the expense of considerable complexity, and they usually ignore the effects of the non-Gaussianity of the density distribution. Our principal goal is to quantify the nonlinear overdensity response for finite volumes beyond linear and second-order SPT to show the effect of the density field distribution and to provide an easy-to-use yet accurate fit as a function of cosmological parameters. This opens the road toward the precise measurement and interpretation of the power spectrum from smaller surveys and simulation subsamples when the overdensity is known from a larger lower-resolution survey or simulation, respectively. In particular, zoom-in simulations \citep{1993ApJ...412..455K, 2014MNRAS.437.1894O}, as well as the recent compactified multi-resolution simulations \citep{2018MNRAS.477.1949R}, will benefit from our results.

The outline of the paper is as follows: first, we describe the density distribution of the cosmic density field. Then, we define the local power spectrum and show the connection between the distribution of the density field and these responses. In Sect. 4 we measure the responses in sub-volumes of larger cosmological $N$-body simulations and give a new fit for these responses that is consistent with the lognormal density field. Finally, we summarize our results.

\section{Density field distribution}

The statistical properties of the $\delta = \rho/\overline{\rho}-1$ Eulerian overdensity field have important effects on the finite volume two-point statistics. The probability distribution function (PDF) of the $\delta$ field is well approximated with the
\begin{equation}
    f_{\text{G}}(\delta,\sigma) = \frac{1}{\sigma\sqrt{2\pi}}e^{-\frac{1}{2}\left(\frac{\delta}{\sigma}\right)^2}
\end{equation}
Gaussian distribution for large scales (e.g. $L>300Mpc$ in standard $\Lambda$ cold dark matter cosmology at $z=0$). For smaller scales, the
\begin{equation}
    f_{\text{ln}}(\delta,\sigma) = \frac{1}{(\delta+1)}\frac{1}{\sigma\sqrt{2\pi}}\cdot e^{-\frac{1}{2}\left(\frac{ln(\delta+1)+\frac{1}{2}\sigma^2}{\sigma}\right)^2}
    \label{eq:lognormalPDF}
\end{equation}
lognormal distribution fits the simulated and observed density fields better \citep{1991MNRAS.248....1C, 2018MNRAS.473.3598R}. The lognormal assumption extends smoothly into the Gaussian regime. The variance, $\sigma^2$, can be calculated from the cosmological parameters for a given volume by
\begin{equation}
    \sigma^2(V) = \int\limits_0^\infty \frac{dk}{2\pi^2} k^2 W(R,k) P(k),
\end{equation}
where $P(k)$ is the power spectrum determined by the cosmological parameters, and $W(R,k)$ is the Fourier representation of a spherical top-hat window function with $R = \sqrt[\leftroot{-2}\uproot{2}3]{3V/(4\pi)}$ radius. This $\sigma^2$ also can be determined from cosmological simulations by dividing the simulation cube into small cubic sub-volumes with $L$ linear sizes and using the
\begin{equation}
    \sigma^2(V) = \frac{1}{N}\sum\limits_{i=0}^N\delta_{i}^2
\end{equation}
formula, where $\delta_i$ is the overdensity of the sub-volume with index $i$, $V=L^3$ is the volume, and $N$ is the total number of the sub-volumes. We note that this $\sigma^2$ is expected to be different from the $\sigma^2_{lin}$ linear mass variance.
The $f_{ln}(\delta,\sigma)$ is more realistic than the Gaussian assumption in that it only assigns positive probability for nonnegative $\delta+1$ densities.

\begin{table}
        \centering
\begin{tabular}{  l | c | c  }
\hline
            Simulation & EdS\_1260 & LCDM\_1260 \\ \hline\hline
        $\Omega_m$ & $1.0$ & $0.3089$ \\
        $\Omega_\Lambda$ & $0.0$ & $0.6911$ \\
            $\Omega_k$ & \multicolumn{2}{c}{$0.0$} \\
        $H_0 \left[\textnormal{km/s/Mpc}\right]$ &  \multicolumn{2}{c}{$67.74$}\\
        $\sigma_8$ & \multicolumn{2}{c}{$0.8159$}\\
            $z_{\textnormal{initial}}$ & \multicolumn{2}{c}{$63$}\\
        \hline
        $N_{\text{part}}$ & $3.43 \cdot 10^8$ & $1.0 \cdot 10^9$\\
        $L_{\text{box}}\left[\textnormal{Mpc}h^{-1}\right]$ & \multicolumn{2}{c}{$1260.0$} \\
        \hline
\end{tabular}
        \caption{Cosmological parameters of the simulations. The $\Lambda$CDM parameters are based on the Planck 2015 results.}
\label{tab:simulations}
\end{table}

\begin{figure}
    \centering
        \includegraphics[width=0.50\textwidth]{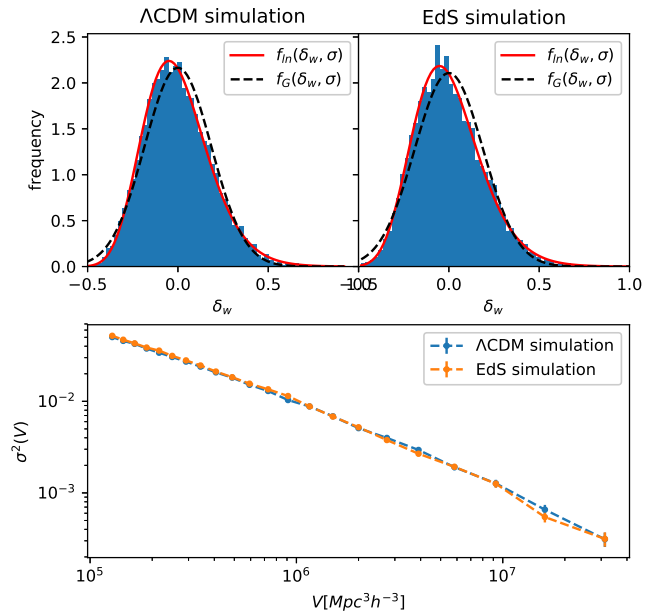}
        \caption{Statistical properties of the simulated cosmic overdensity field. \textbf{Top:} Measured distribution of the $\delta_w$ overdensity field and the $f_{ln}(\delta_w,\sigma)$ lognormal distribution function for $L=60\text{Mpc}h^{-1}$ cubic windows in our $\Lambda$CDM and EdS simulations at $z=0$ redshift. We plotted the Gaussian approximation of the density field as a dashed black curve. \textbf{Bottom:} Measured $\sigma^2$ variance of the $\delta_w$ field as a function of the window volume at $z=0$.}
        \label{fig:sigmaV}
\end{figure}

We used Einstein-de Sitter (EdS) and standard $\Lambda$ cold dark matter ($\Lambda$CDM) cosmological $N$-body simulations to calculate $\sigma(V)$ at redshift $z=0$. The $\Lambda$CDM simulation had cosmological parameters taken from  \cite{2016A&A...594A..13P}, and all initial conditions were generated with the 2LPTic code \citep{2012ascl.soft01005C, 2006MNRAS.373..369C}, with initial redshift $z_{\text{start}}=63$. The simulations were done with the GADGET-2 code \citep{2005MNRAS.364.1105S} (see Table~\ref{tab:simulations} for parameters).  Figure~\ref{fig:sigmaV} displays the derived density distributions and the $\sigma^2(V)$.


\section{The power spectrum within finite volumes}

The power spectrum of the $\delta(\mathbf{x})$ overdensity field is defined by 
\begin{equation}
    (2\pi)^3P(\mathbf{k})\delta^3_D(\mathbf{k}-\mathbf{k}') = \left\langle\delta(\mathbf{k})\delta(\mathbf{k}') \right\rangle,
\end{equation}
 where $\delta(\mathbf{k})$ is the Fourier transform of the $\delta(\mathbf{x})$ field and $\delta^3_D(\mathbf{k})$ is the three-dimensional Dirac-delta function. This function contains all information from the statistics of the density field in the linear regime, and a decreasing fraction of the total information on smaller scales, if initial density fields followed Gaussian statistics.  Since we assume that the Universe is isotropic, $P(\mathbf{k})$ depends only on the length of the $\mathbf{k}$ vector and is thus denoted as $P(k)$.


We adapted the definition by \citet{Chiang_2014} for the position-dependent power spectrum. For the rest of the paper we use the cubic window function 
\begin{equation}
    W_L(\mathbf{x}) = \prod\limits_{i=1}^{3}\theta(x_i), \; \text{where} \; \theta(x_i) = \left\{
                \begin{array}{l l}
                1 & \text{if } |x_i|<L/2 \\
                0 & \text{otherwise}
                \end{array} \right. 
    \label{eq:windowfunction}
,\end{equation}
 where $L$ is the linear size and $\mathbf{x}$ is a comoving coordinate vector. The local Fourier transform of the density field in this case is
\begin{equation}
    \delta(\mathbf{k},\mathbf{r}_w,L) = \int d^3x \delta(\mathbf{x})W_L(\mathbf{x}-\mathbf{r}_w)e^{-i\mathbf{k}\mathbf{x}},
\end{equation}
where $\mathbf{r}_w$ is the comoving coordinate of the center of the survey. The position-dependent power spectrum inside the survey volume is then constructed as
\begin{equation}
    P_w(k,\mathbf{r}_{w},L) =  \frac{1}{V}\left\langle|\delta(\mathbf{k},\mathbf{r}_w,L)|^2\right\rangle,
\end{equation}
where $V=L^3$ is the volume of the cubic survey. The local power spectrum definition above assumes that the mean cosmic mass density is known for the power spectrum analysis. This is true for simulations, but not for real surveys. If the global density is unknown, it is estimated from the average density inside the survey window, and the
\begin{equation}
\tilde{\delta}(k) = \frac{\delta(k)}{1+\delta_w}    
\end{equation}
field is used in the power spectrum calculation instead of $\delta(k)$
\citep{2012JCAP...04..019D}. This causes a bias in the power spectrum estimation, and
the power spectrum in the survey window becomes
\begin{equation}
    P_{sw}(k) = \frac{1}{(1+\delta_w)^2}P_w(k)
    \label{eq:surveylocalpk}
,\end{equation}
where $P_w(k)$ is the local power spectrum defined above. This effect on the measurements is called the local average effect. Since we work with simulations in this paper, we used the $P_w(k)$ local power spectrum calculated by the known $\overline{\rho}$ average density unless otherwise stated.

The power spectrum and the position-dependent power spectrum evolves over time. There are two commonly used methods available to predict the power spectrum from an initial state at a later time: PT and numerical simulations of structure formation.
The former are accurate early on or on the largest scales for late cosmological times, when structure formation is linear or mildly nonlinear. In the nonlinear regime, only the numerical simulations and the halo model yield precise predictions.

The evolution of the local power spectrum at coordinate  $\mathbf{r}_w$ within  volume $V$ is determined by the cosmological parameters, the initial fluctuations, the overdensity inside the window ($\delta_w$), and the tidal fields originating from outside the survey area. It is a difficult task to calculate these effects properly in the nonlinear regime. For simplicity, we neglected tidal fields and modeled the ratio of the position-dependent power spectrum and the global power spectrum as
\begin{equation}
    P_w(k,t|\delta_w,\sigma(V)) = R(k,t,\delta_w,\sigma(V))P(k,t),
    \label{eq:GeneralPkResponse}
\end{equation}
where $\sigma^2(V)$ is the variance of the overdensity field on the scale of the window function and $R(k,t,\delta_w,\sigma(V))$ is the {response function}. It quantifies the response of the position-dependent power spectrum  to the presence of a large-scale overdensity ($\delta_w$) inside the window \citep{ Chiang_2014, 2015JCAP...08..042W}.

By definition, the position-dependent power spectrum for every sub-volume, $V$, should average  -- after deconvolving the window function -- to the global power spectrum. Thus, the response function fulfills
\begin{equation}
    1 = \frac{\left\langle P_w(k,t|\delta_w,V) \right\rangle}{P(k,t)} = \int  R(k,t,\delta_w,\sigma(V)) f(\delta_w,\sigma(V)) d\delta_w,
    \label{eq:ResponseAverage}
\end{equation}
 where $f(\delta_w,\sigma(V))$ is the probability density-distribution function of the overdensity field on the scale of the window function.

\subsection{Perturbative power spectrum responses}

In SPT, the effect of the overdensities are described as a function of the $\delta_{L0}D(t)$ {linearly extrapolated} Lagrangian overdensity,
\begin{equation}
    P(k,t|D(t)\delta_{L0}) = \sum\limits_{n=0}^{\infty} \frac{1}{n!} R_{\mathcal{L},n}(k,t)\left[\delta_{L0}D(t)\right]^n \overline{P}(k,t),
    \label{eq:PkExpansion}
\end{equation}
where  $\delta_{L0}$ is the {initial} overdensity, $D(t)$ is a linear growth function, and
\begin{equation}
    R_{\mathcal{L},n}(k,t) = \left.  \frac{1}{\overline{P}(k,t)} \frac{d^n P(k,t|\delta_{L0}D(t)) }{d\left(\delta_{L0}D(t)\right)^n}\right|_{\delta_{L0}D(t)=0}
    \label{eq:PerturbativeResponses}
\end{equation}
is the $n$th-order response function, with the zeroth order set to $R_{\mathcal{L},0}(k,t)=1$ \citep{2015JCAP...08..042W}. An initially Gaussian field is assumed, and therefore the distribution of the linearly extrapolated density field is Gaussian too. Each power of the extrapolated overdensity is zero for a sub-volume with an average density. The consequence of this fact is that any order SPT response that uses this Taylor expansion will predict a power spectrum that is identical to the global one for average density sub-volumes.


The first two orders of the response function are the following:
\begin{equation}
    R_{\mathcal{L},1} = \frac{47}{21} - \frac{1}{3} \frac{d \ln P(k)}{d \ln k}
\end{equation}
\begin{equation}
    R_{\mathcal{L},2} = \frac{8420}{1323} - \frac{100}{63}\frac{k}{P(k)}\frac{d P(k)}{dk} + \frac{1}{9}\frac{k^2}{P(k)^2}\frac{d^2P(k)}{dk^2}
.\end{equation}
The Eulerian responses can be calculated for the EdS Universe as\begin{equation}
    R_1(k) = R_{\mathcal{L},1}(k)
\end{equation}
\begin{equation}
    R_2(k) = R_{\mathcal{L},2}(k) - \frac{34}{21}R_{\mathcal{L},1}(k)
\end{equation}
\citep{2015JCAP...08..042W}. As we show next, these results fit cosmological simulations well on scales where the probability distribution of the overdensity, $\delta$, is nearly Gaussian.

The response function constructed from $R_0$ and $R_1$  fulfills Eq.~\ref{eq:ResponseAverage} for a Gaussian distribution. While this is no longer true for higher-order SPT,  the average should converge to one with increasing orders.


\section{Response function from cosmological simulations}

We expect the PT responses to accurately predict the  position-dependent power spectra when $\delta_w$ is sufficiently close to zero and the density field is close to a Gaussian. If the survey window is too small, the skewness of the probability density function distorts the responses. In particular, the power spectrum of the average density sub-volumes might differ from the spectrum of the full cosmological volume due to non-Gaussianity.

\begin{figure}
    \centering
        \includegraphics[width=0.50\textwidth]{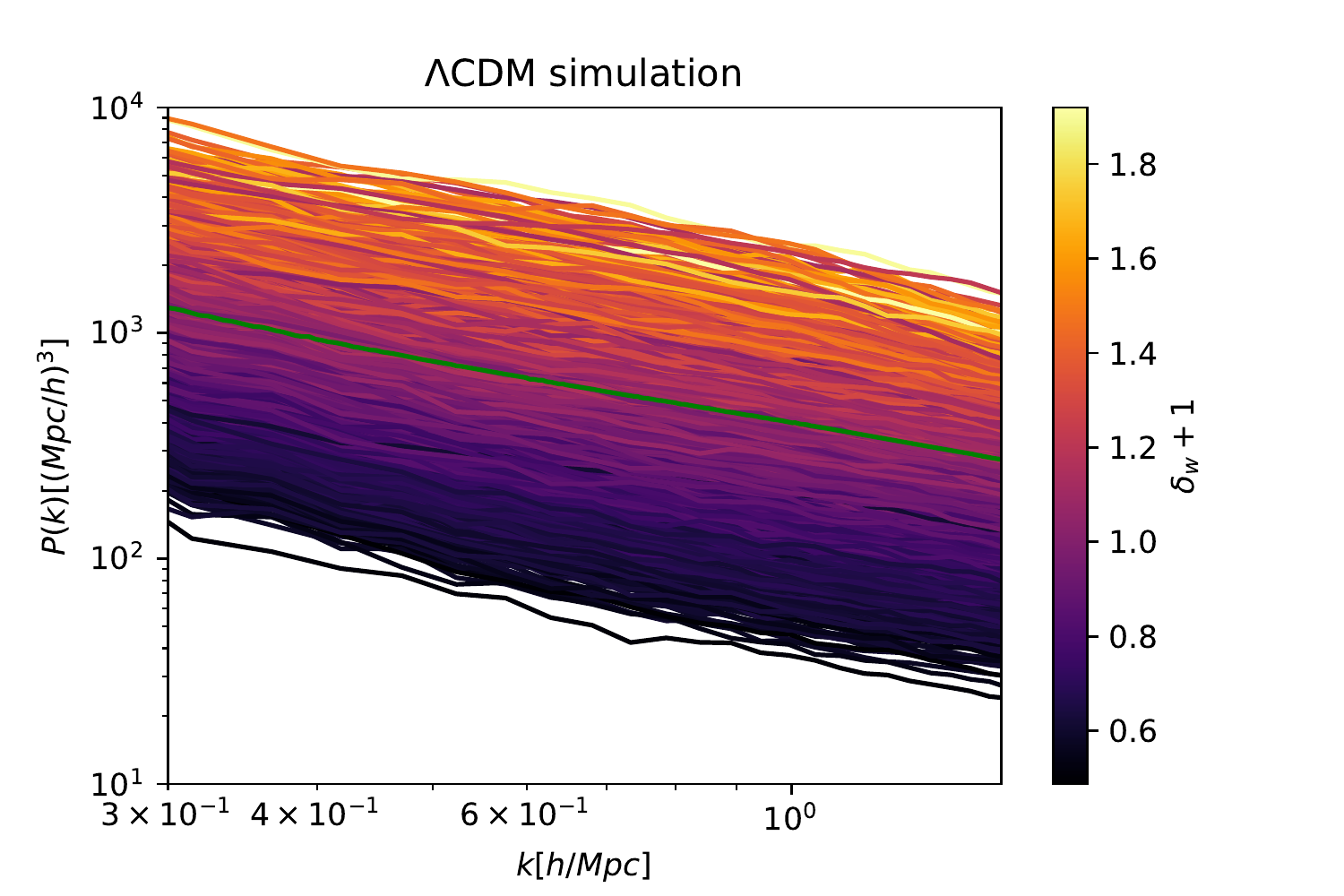}
        \caption{Position-dependent power spectra in our $\Lambda$CDM simulation at $z=0$ with a $L=60\text{Mpc}h^{-1}$ window size. The color represents the average density of each sub-volume. The full-volume power spectrum is plotted as a green curve for reference.}
        \label{fig:LocalPks}
\end{figure}

To test this hypothesis, we calculated the position-dependent power spectra in a large number of sub-volumes from the simulations shown earlier. We divided the simulation volume into $N_{cut}^3$ distinct cubic sub-volumes with $L=L_{box}/N_{cut}$ side lengths, similarly  to \citet{Chiang_2014}. The position-dependent power spectra and the full-volume spectrum for $L=60\text{Mpc}h^{-1}$ in a $\Lambda$CDM simulation is shown in Fig.~\ref{fig:LocalPks}.

\begin{figure}
    \centering
        \includegraphics[width=0.50\textwidth]{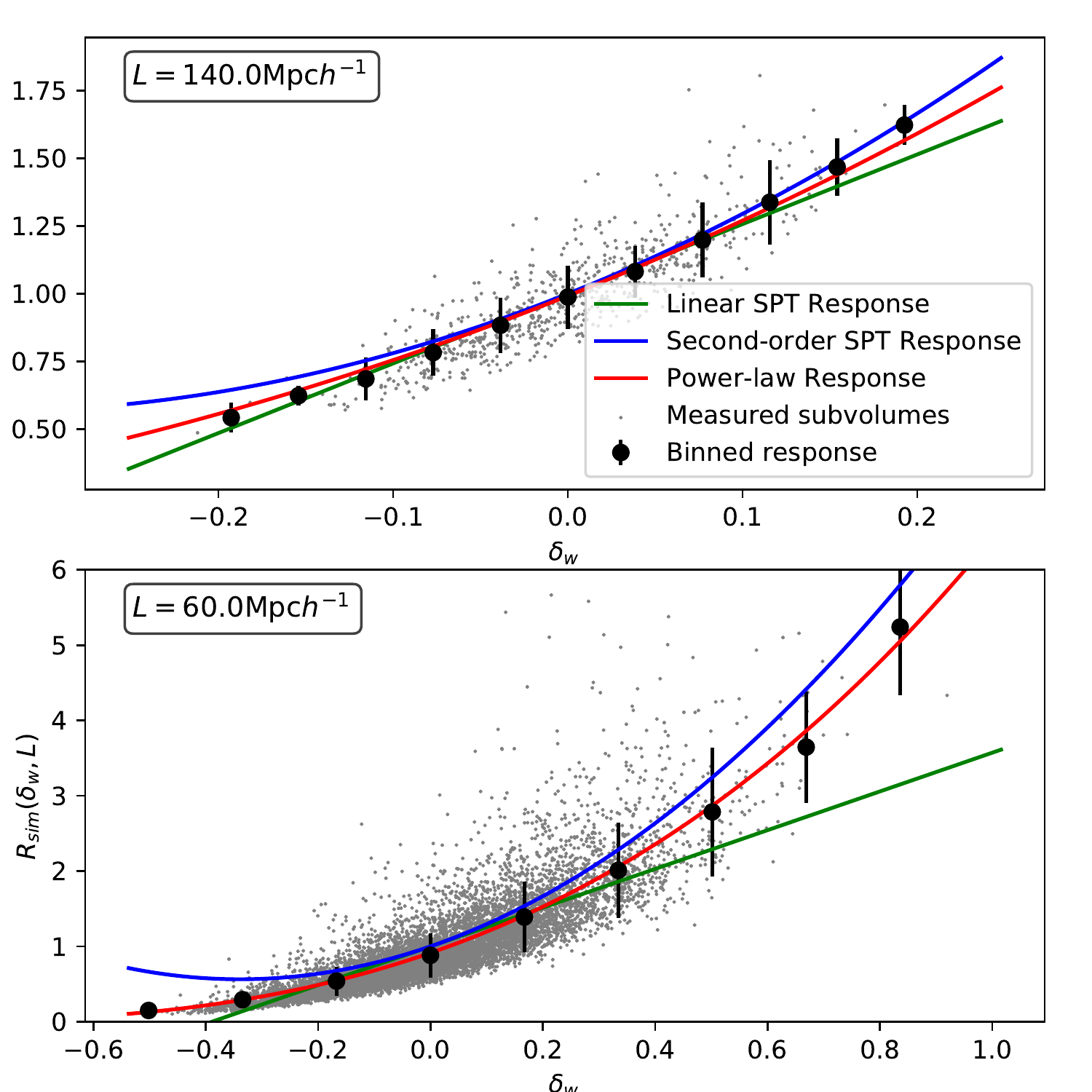}
        \caption{Simulated responses from our $\Lambda$CDM simulation with two different $L$ window sizes at $z=0$. The small gray markers represent the measured sub-volumes. This data have been re-binned, and the black circles with error bars represent the average response in each bin and the $1\sigma$ deviation. The linear ($R=1+R_1\delta_w$) and second-order ($R=1+R_1\delta_w+R_2\delta_w^2$) responses from SPT are plotted with the green and blue curves, respectively. The red curves represent the new, $\sigma$-dependent power-law response.}
        \label{fig:SimulatedResponses}
\end{figure}

Compared to the global spectrum, the position-dependent spectra are only shifted by a constant factor for the $k>0.3\text{Mpc}^{-1}h$ region, suggesting that the overdensity response function is at most extremely weakly dependent on the $k$ wavenumber. Motivated by this, we neglected any $k$ dependence and defined the response for the $i$-th sub-volume as
\begin{equation}
    \begin{split}
    R_{\text{sim, }i}(\delta_w,L,t) &= \left\langle\frac{P_{w,i}(k,\delta_w,L,t)}{P(k,t)}\right\rangle_{k_{min}<k<k_{max}}=\\
    &=\frac{1}{k_{max}-k_{min}}\int\limits_{k_{min}}^{k_{max}}dk\frac{P_{w,i}(k,\delta_w,L,t)}{P(k,t)},
    \end{split}
    \label{eq:SimulatedResponse}
\end{equation}
where $\langle\rangle_k$ denotes $k$-average. Since the position-dependent power spectrum is sampled on discrete $k_j$ values in our case, Eq.~\ref{eq:SimulatedResponse} becomes
\begin{equation}
    R_{\text{sim, }i}(\delta_w,L,t) = \frac{1}{N_k}\sum\limits_{k_{min}<k_j<k_{max}} \frac{P_{w,i}(k_j,\delta_w,L,t)}{P(k_j,t)},
    \label{eq:SimulatedResponseNumeric}
\end{equation}
where $N_k$ is the number of $k_j$ modes that satisfy $k_{min}<k_j<k_{max}$. The $k_{min}$ and $k_{max}$ values were set to $\textnormal{max}\left(16\cdot\pi/L, 0.3 \text{Mpc}^{-1}h\right)$ and $1.5\text{Mpc}^{-1}h $, respectively, to minimize the effect of the window function and the discreteness of the particles. The measured $R_{\text{sim, }i}(\delta_w,L,t)$ is shown along with the SPT responses in Fig.~\ref{fig:SimulatedResponses}. The first- and second-order perturbative responses were calculated by assuming a power law for P(k), and by calculating $k$-average between $k_{min}$ and $k_{max}$. The power-law approximations were valid in all measured $k$ ranges. According to a visual inspection, the SPT response is consistent with the measured one in the larger windows when the overdensity, $\delta_w$, is not far from zero. In smaller windows, especially for negative overdensities, SPT is far off, not even obeying the positivity constraint. Motivated by Fig.~\ref{fig:SimulatedResponses}, we propose a new power-law fit for the response function,
\begin{equation}
    R(\delta_w,\sigma) = B(\sigma)\left(\delta_w+1\right)^{A(\sigma)}.
    \label{eq:PowerLawResponse}
\end{equation}
Since our goal is to predict the response from cosmological parameters, we express the dependence on the scale $L$ as a function of $\sigma$ dependence since $\sigma(V) = \sigma(L^3)$ is a known bijective function given a cosmological model.
The response in this form is always positive when $\delta_w > -1$, and it fits the simulated responses at all scales and overdensities. Our next objective was to calculate the $B(\sigma)$ and $A(\sigma)$ functions.

We required our fit to be compatible with the linear-order responses from SPT for low $\delta_w$ overdensity and large scales. Thus,
\begin{equation}
    \left. \frac{d R(\delta_w,\sigma)}{d\delta_w}\right|_{\delta_w=0} = R_{1} = \frac{47}{21}-\frac{1}{3}\frac{d \ln P(k)}{d \ln k}.
    \label{eq:ResponseLinearConnection}
\end{equation}
Therefore, $B(\sigma)$ and $A(\sigma)$ are not independent, and Eq.~\ref{eq:PowerLawResponse} takes the
\begin{equation}
    R(\delta_w,\sigma) = B(\sigma)\left(\delta_w+1\right)^{R_1/B(\sigma)}
    \label{eq:PowerLawResponseB}
\end{equation}
form. For a late time $\Lambda$CDM cosmology beyond the weak linear regime,  $\frac{d \ln P(k)}{d \ln k}$ is well approximated by $-1$ \citep{2013MNRAS.434.2961C}.  

\begin{figure*}
  \centering
        \includegraphics[width=0.95\textwidth]{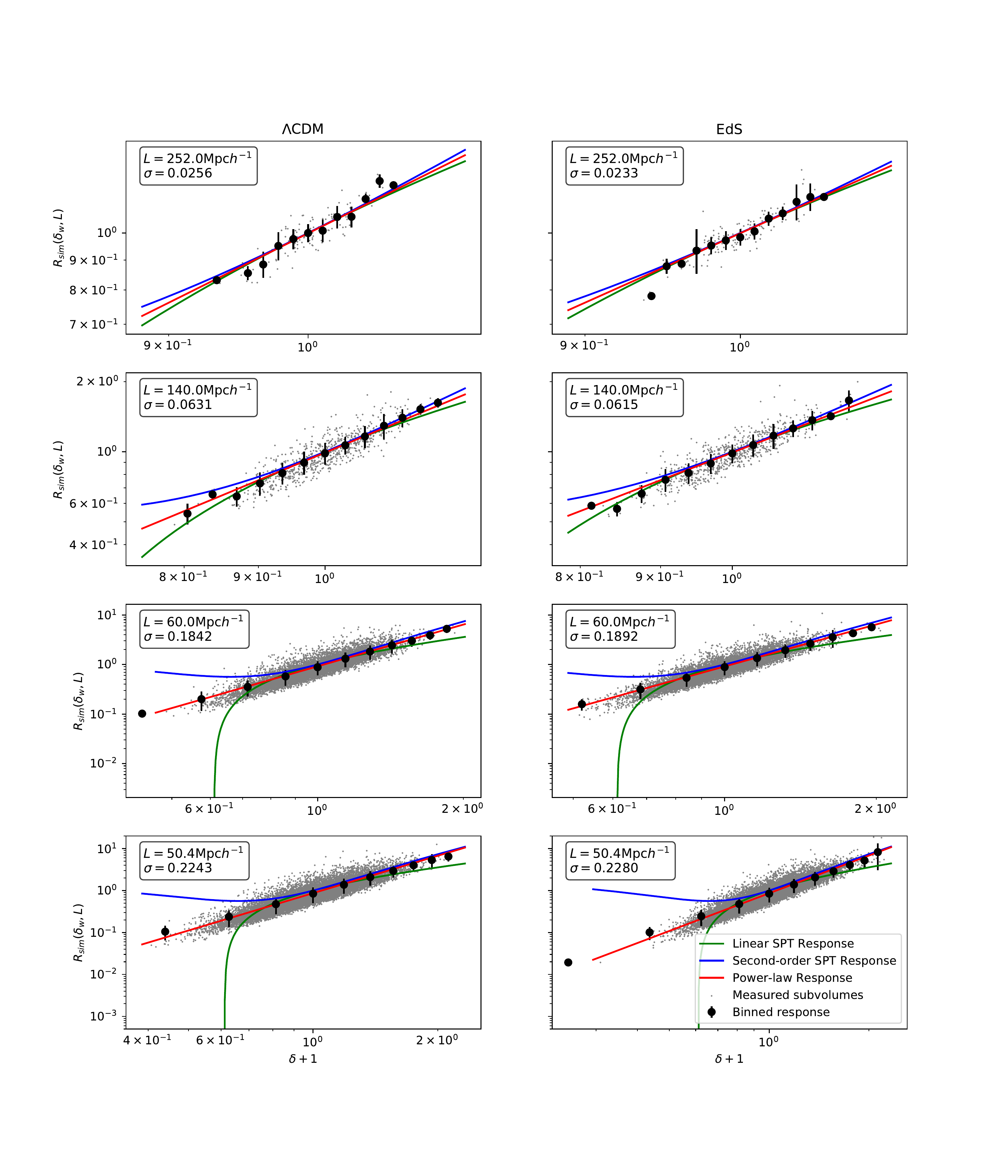}
        \caption{Simulated and theoretical responses in different window scales for two different cosmological models at $z=0$. The new power-law response function fits the simulated responses well. \textbf{Left:} $\Lambda$CDM cosmology. \textbf{Right:} EdS cosmology.}
        \label{fig:PowerLawResponse}
\end{figure*}

The sub-volume power spectra should average to the global power spectrum. Using the lognormal probability density function for $\delta_w$, the consistency requirement of Eq.~\ref{eq:ResponseAverage} for our response function becomes 
\begin{equation}
    B(\sigma)\cdot e^{\frac{\sigma^2}{2}\left[\left(\frac{R_1}{B(\sigma)}\right)^2-\frac{R_1}{B(\sigma)}\right]} = 1.
    \label{eq:biasequation}
\end{equation}
This equation determines the $B(\sigma)$ function. This is a transcendental equation with no known analytical solution. We used numerical solutions to calculate the responses in Figs.~\ref{fig:SimulatedResponses} and~\ref{fig:PowerLawResponse}. Approximate solutions can be derived from Taylor series. The detailed calculations are in Appendix~\ref{sec:AppA}. The first-order approximate solution suffices for most applications:
\begin{equation}
    B_{I}(\sigma) = 1 - \frac{1-e^{-\frac{1}{2}\sigma^2\left(R_1^2-R_1\right)}}{\left(\frac{1}{2}\sigma^2\left(R_1-2R_1^2\right) + 1\right) }.
    \label{eq:AsigmaFirstOrderSolution}
\end{equation}
The error of this approximation is below one percent for scales where $\sigma<0.24$. This solution can be further approximated by expanding it as
\begin{equation}
    B(\sigma) \simeq 1 - \frac{1}{2}\sigma^2R_1\left(R_1-1\right) \simeq 1-2\sigma^2 \\ (\text{for}\\ \frac{d \ln P(k)}{d \ln k}\simeq -1).
    \label{eq:Asigma2ApproximateSolution}
\end{equation}
This is accurate for window sizes with $\sigma<0.1$ with sub-percent error.
The median for the log-normal density field defined in Eq.~\ref{eq:lognormalPDF} is
\begin{equation}
    \textnormal{median}(\delta_w) = e^{-\frac{1}{2}\sigma^2} - 1.
\end{equation}
Using the Taylor expansion of Eqs. 27 and Eq.~\ref{eq:Asigma2ApproximateSolution}, the connection between the super survey bias and the median of the overdensity field can be written as
\begin{equation}
     B \simeq 1 + 4\cdot \textnormal{median}(\delta_w).
\end{equation}

\begin{figure}
  \centering
        \includegraphics[width=0.45\textwidth]{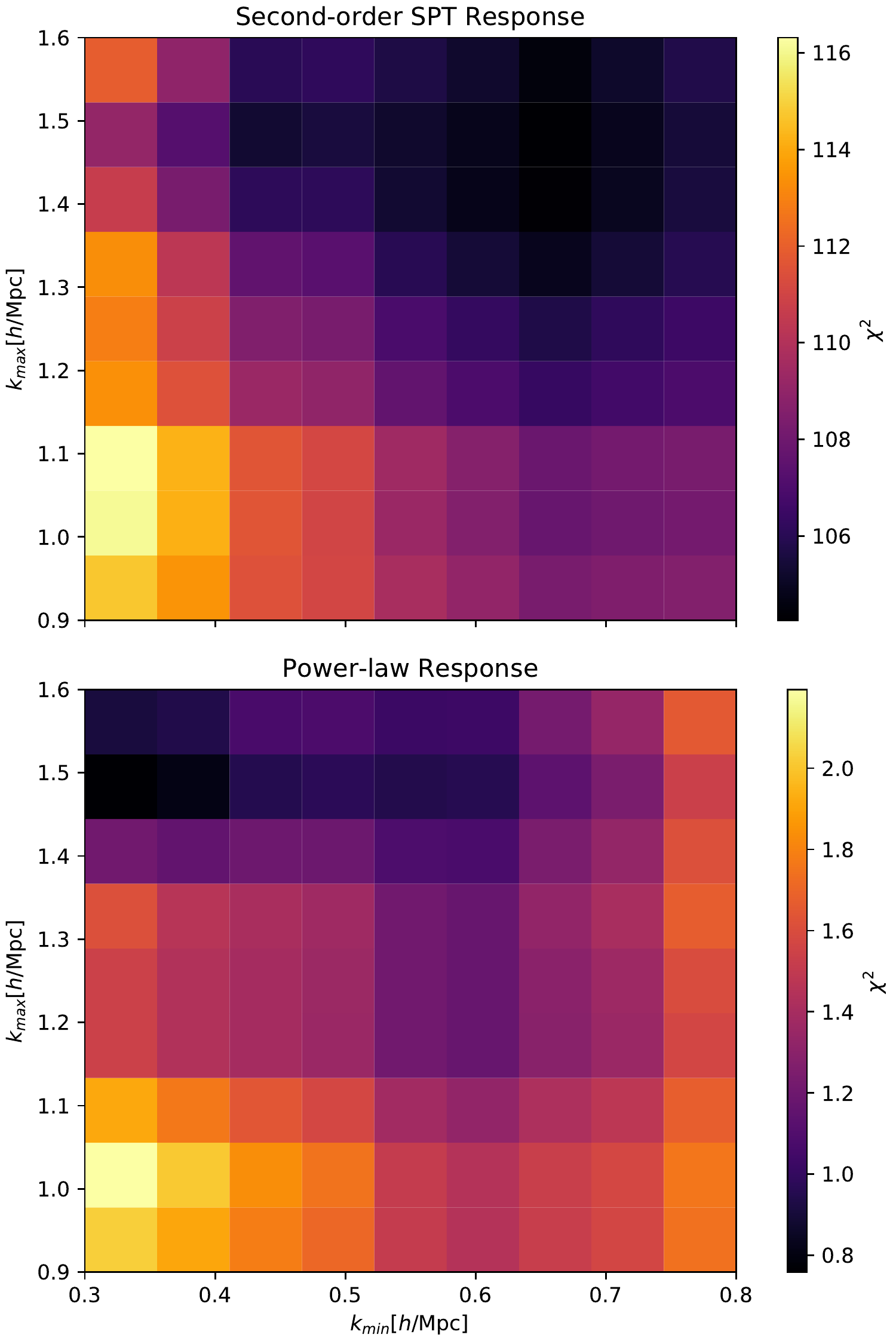}
        \caption{Comparison of the goodness of the fit of the theoretical responses in the $L=60\text{Mpc}h^{-1}$ window size at $z=0$. The new power-law response function fits the simulated responses significantly better.}
        \label{fig:kminkmaxChi2}
\end{figure}

To test the effect of the $k_{min}$ and $k_{max}$ in Eq.~\ref{eq:SimulatedResponseNumeric}, we binned the calculated responses into $N_b=13$ equally spaced $\delta_w$ bins for multiple $(k_{min},k_{max})$ pairs and calculated the
\begin{equation}
     \chi^2(k_{min},k_{max}) = \sum\limits_{i=0}^{N_b}\frac{\left(R_{theory}(\delta_i)-R_{sim,i}(k_{min},k_{max})\right)^2}{\sigma^2_{sim,i}(k_{min},k_{max})}
\end{equation}
quantity for the second-order SPT and the new power-law responses, where $R_{sim,i}$ is the average response inside the $\delta_i$ bin and $\sigma^2_{sim,i}$ is the variance of the simulated response. The calculated $\chi^2(k_{min},k_{max})$ showed that our power-law response fits the simulated responses significantly better compared to the second-order SPT response, and the goodness of this fit only mildly depends on the chosen $(k_{min},k_{max})$ pairs. This can be seen in Fig.~\ref{fig:kminkmaxChi2}.

Our proposed power-law fit and its approximations work in all regimes, even where the SPT approximation breaks down, despite having the same number of parameters as the second-order SPT. While SPT does a good job in both $\Lambda$CDM and in EdS cosmology for high density sub-volumes, it fails catastrophically for low density regions, for the first and second order. These are most of the sub-volumes due to the lognormal distribution of overdensities, and this is exactly where our proposed fit works significantly better than any previous approach. Higher-order SPT responses can achieve better fits \citep{2015MNRAS.448L..11W}, but they are significantly more complicated compared to this new formula.

\begin{figure}
  \centering

        \centering
        \includegraphics[width=0.45\textwidth]{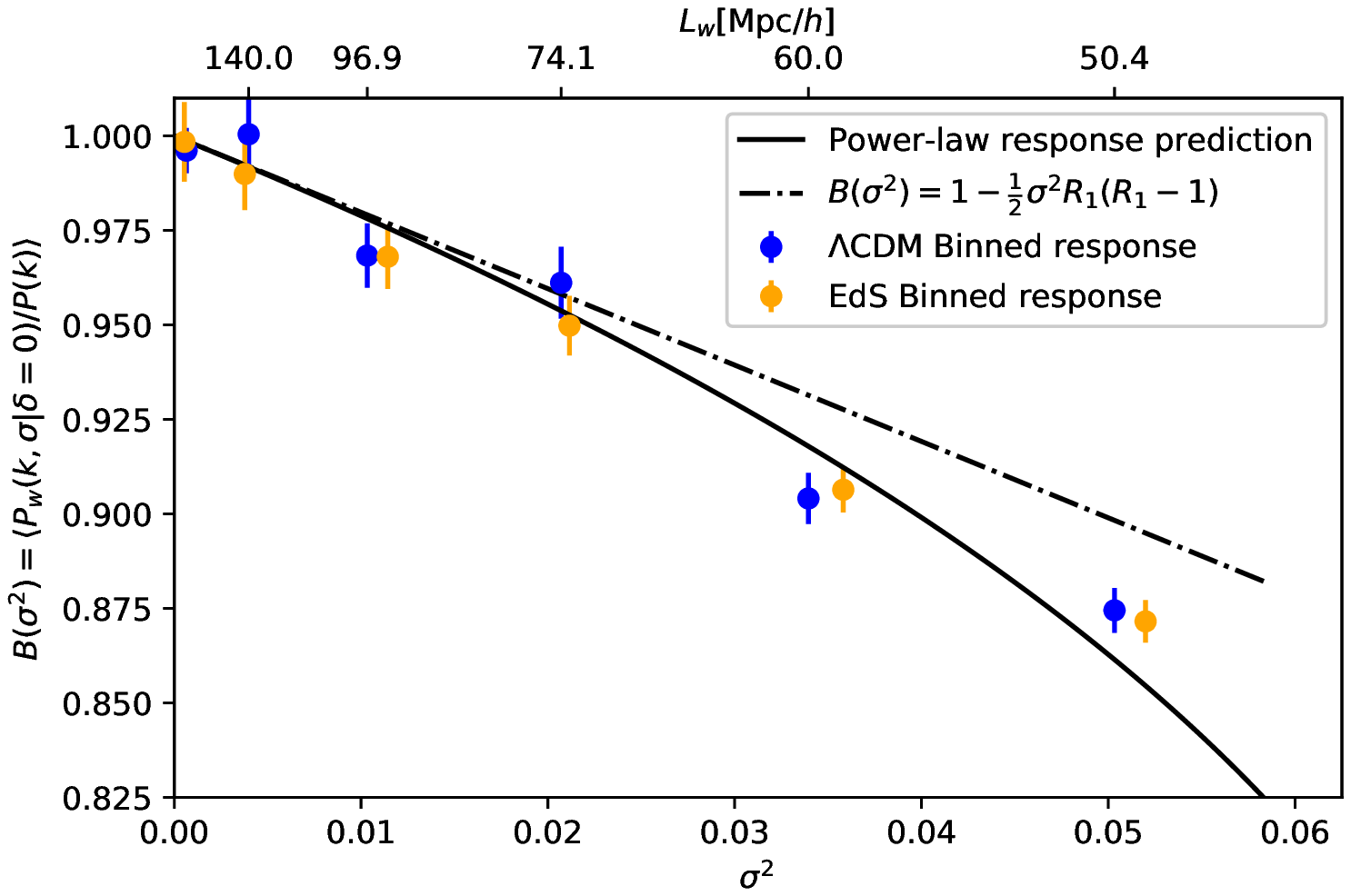}
        \centering
        \includegraphics[width=0.45\textwidth]{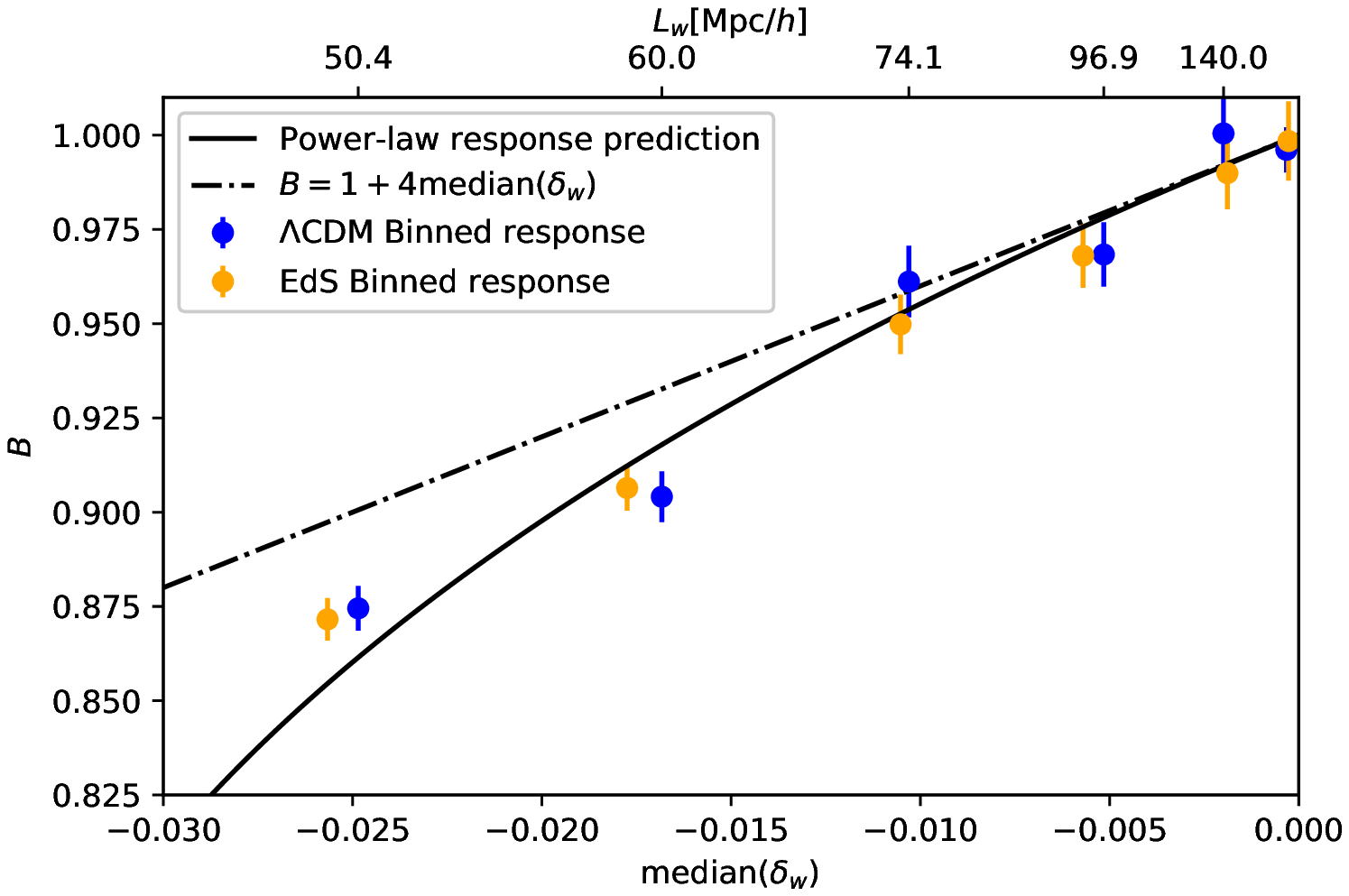}

        \caption{Simulated $R_{sim}(\delta_w=0,\sigma)$ responses for $\Lambda$CDM and EdS cosmology. In contrast to SPT, the power-law response predicts that the mean density sub-volumes have smaller than average power spectra. This numeric $B(\sigma^{2})$ prediction is plotted with a black curve and shows good agreement with the simulated responses in both cosmologies. \textbf{Top:} Measured and predicted bias as a function of $\sigma^2$. \textbf{Bottom:} Bias as a function of the median sub-volume overdensity, $e^{-\frac{1}{2}\sigma^2}-1$.}
        \label{fig:AsigmaLCDMEdS}
\end{figure}

Another physical consequence of the lognormal density field distribution and the power-law response is the bias of the power spectrum in the average density sub-volumes. According to Eqs.~\ref{eq:GeneralPkResponse} and~\ref{eq:PowerLawResponse}, the $\delta=0$ volumes on average have $P_w(k,t|\delta_w=0,\sigma) = B(\sigma)P(k,t)$ as opposed to SPT, where these regions have the same spectrum as the full-volume one. To test this this prediction, we plotted the simulated responses of the $\delta_w=0$ bins from Fig.~\ref{fig:PowerLawResponse} as a function of the $\sigma$ deviation with the $B(\sigma)$ responses in Fig.~\ref{fig:AsigmaLCDMEdS}. The simulated $\delta_{w}=0$ bias in the position-dependent power spectrum agrees well with the prediction of the power-law responses in $\Lambda$CDM and EdS cosmology.

\subsection{Redshift dependence}

\begin{figure}
  \centering
        \includegraphics[width=0.50\textwidth]{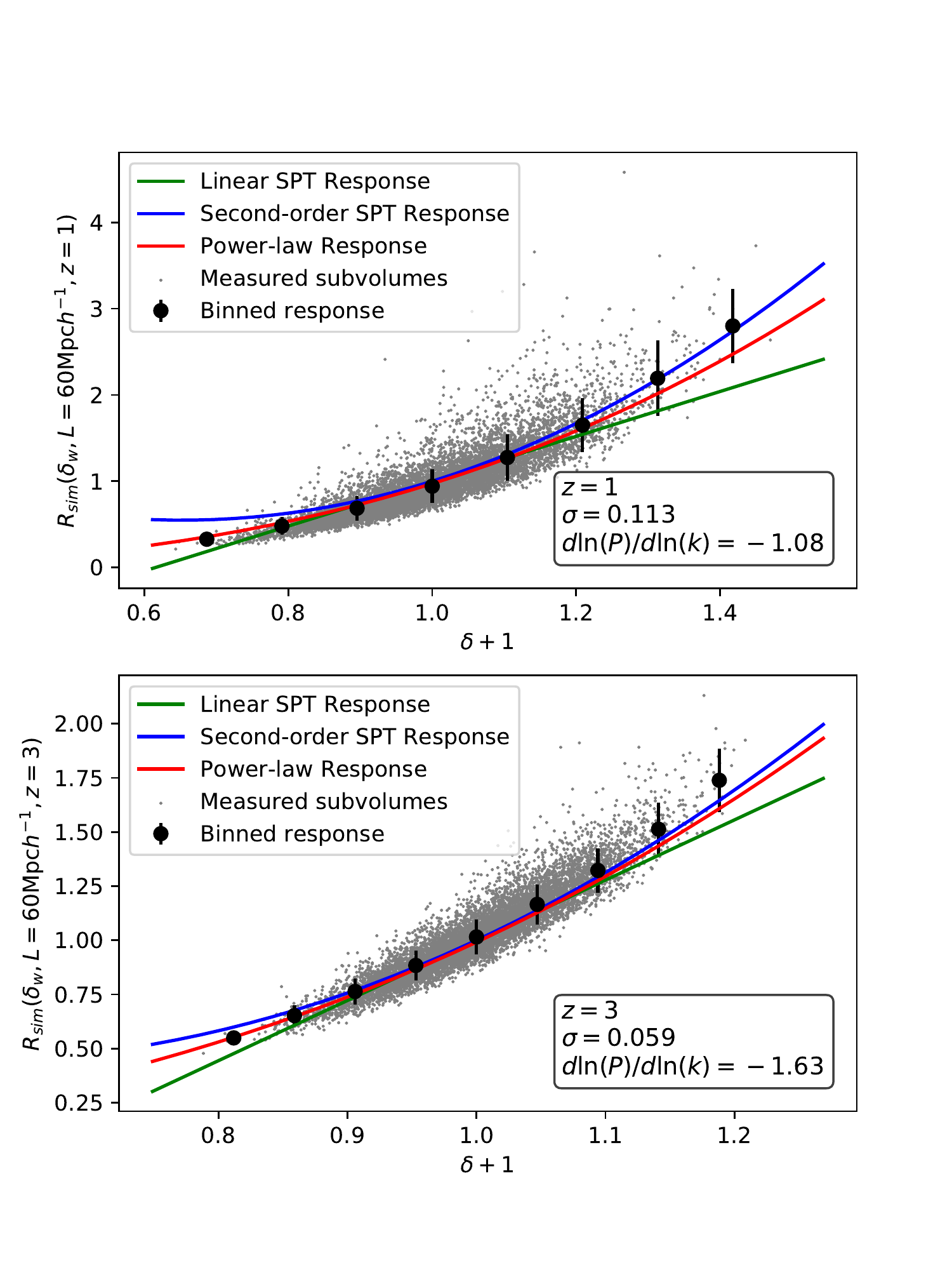}
        \caption{Simulated and theoretical $R_{sim}(\delta_w,\sigma, z)$ responses in $\Lambda$CDM cosmology for the $L = 60\text{Mpc}h^{-1}$ window size at $z=1$ and $3$ redshifts. The new power-law response function fits the simulated responses well, especially in the low density regions. At higher redshifts, the SPT and power-law responses converge as the density field becomes more Gaussian.}
        \label{fig:RedshiftDependency}
\end{figure}

The SPT responses defined in Eq.~\ref{eq:PerturbativeResponses} are dependent on the derivatives of the global power spectrum. As a consequence, these responses depend on the redshift since the power spectrum changes over time. Since we only used the first-order result in our new power-law response, Eq.~\ref{eq:PowerLawResponse} depends only on the redshift through $\frac{d \ln P(k)}{d \ln k}(z)$ and through $\sigma_L(z)$ for a given $L$ scale. We expect that the new power-law response can be universally used for all redshifts if the above redshift dependences  are correctly taken into account. To test this hypothesis, we calculated the responses at $z=1$ and $3$ values for the $L = 60\text{Mpc}h^{-1}$ window size from our $\Lambda$CDM simulation. As can be seen in Fig.~\ref{fig:RedshiftDependency}, the calculated responses agree well with the power-law response function. The $\sigma(z)$ variance of the density field distribution is a monotonically decreasing function of redshift for a given scale. The lognormal density distribution defined in Eq.~\ref{eq:lognormalPDF} converges to Gaussian as the redshift increases and, as a consequence, the difference between perturbative and power-law response predictions decreases.

\subsection{Response inside surveys}

\begin{figure*}
  \centering
        \includegraphics[width=0.95\textwidth]{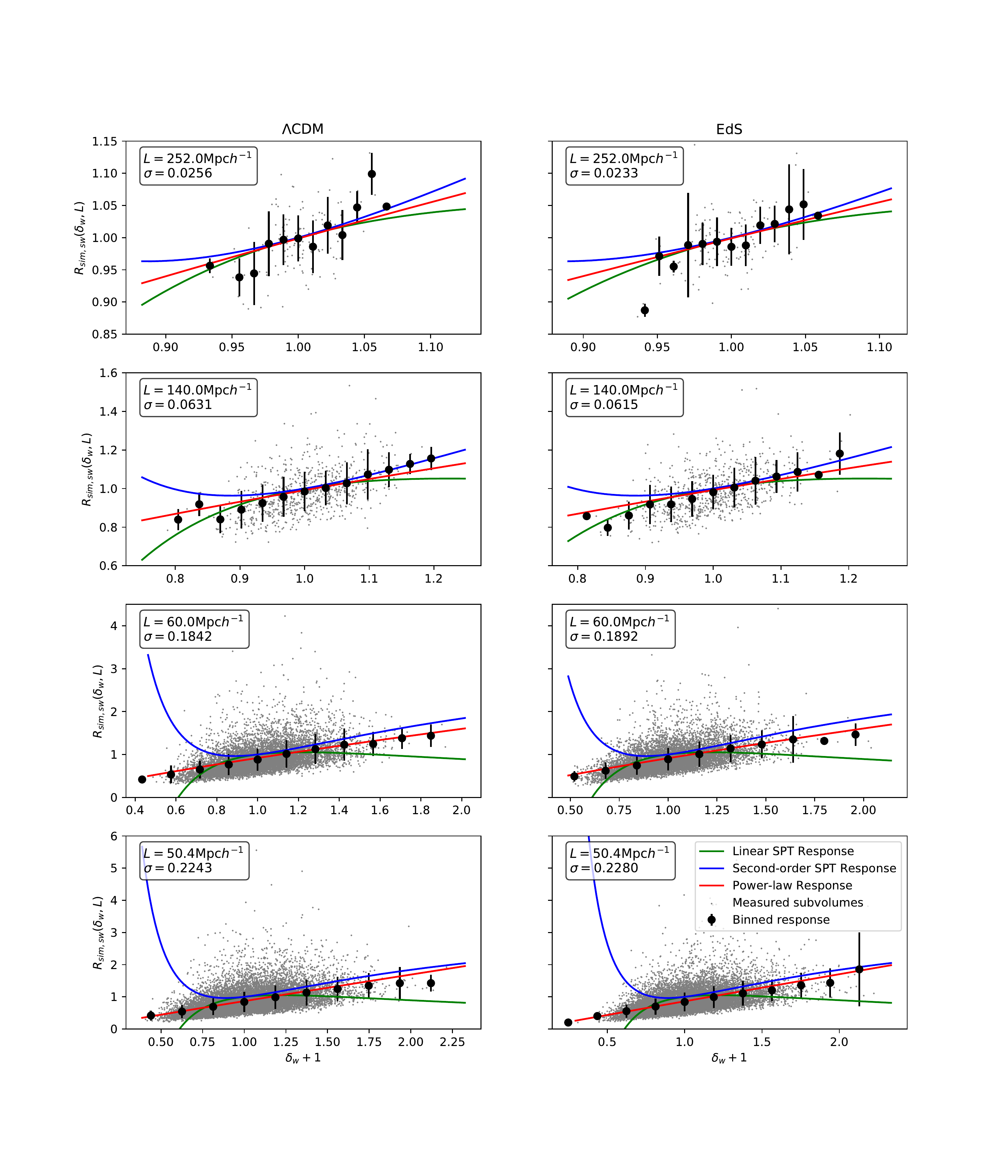}
        \caption{Simulated and theoretical responses in simulated surveys where the average mass density is unknown. The local average effect results in a $(1+\delta_w)^{-2}$ bias in the power spectrum and in the response function compared to Fig.~\ref{fig:PowerLawResponse}. The power-law response in this case is in good agreement with the simulated data. \textbf{Left:} $\Lambda$CDM cosmology. \textbf{Right:} EdS cosmology.}
        \label{fig:SimulatedSurveyResponses}
\end{figure*}

As we stated earlier, there are two ways to calculate the local power spectrum: by using the global average $\overline{\rho}$ mass density, or by using the estimated $\overline{\rho}_{sw} = \overline{\rho} \cdot (1+\delta_w)$ mass density during the power spectrum calculation. The conversion between the two definitions can be done easily by using Eq.~\ref{eq:surveylocalpk}. In the majority of this paper, we used the former method because the global density was available in our simulations. However, for real surveys only the latter definition can be used. We used Eqs.~\ref{eq:surveylocalpk} and \ref{eq:SimulatedResponse} to calculate the
\begin{equation}
    \begin{split}
    R_{sim,sw,i}(\delta_w,L,t) &= \left\langle\frac{P_{sw,i}(k,\delta_w,L,t)}{P(k,t)}\right\rangle_{k_{min}<k<k_{max}}=\\ 
    &= \frac{1}{(1-\delta_w)^2}R_{sim,i}(\delta_w,L,t)
    \end{split}
    \label{eq:SimulatedSurveyResponses}
\end{equation}
simulated survey responses in this case and plotted the results in Fig.~\ref{fig:SimulatedSurveyResponses} for $\Lambda$CDM and EdS cosmology.

The theoretical responses also have to be scaled by $1/(1+\delta_w)^2$ if the power spectrum is calculated from the estimated average mass density, and the power-law response in Eq.~\ref{eq:PowerLawResponseB} thus becomes
\begin{equation}
    R_{sw}(\delta_w,\sigma) = B(\sigma)\left(\delta_w+1\right)^{\left[R_1/B(\sigma)-2\right]}.
    \label{eq:SurverPowerLawResponseB}
\end{equation}
We plot the scaled theoretical responses in Fig.~\ref{fig:SimulatedSurveyResponses}. We note that the bias described by the $B(\sigma)$ function for $\delta_w=0$ average density surveys is also present in this case.

\section{Conclusion}

We have investigated the power spectrum response to overdensities in cosmological $N$-body simulations, paying special attention to the non-Gaussian effects of the density field distribution. Standard perturbation theory  predicts the measured responses on scales where the PDF of the density field is close to a Gaussian distribution. For smaller window sizes, where the distribution is more lognormal, the simulated responses did not match the first- and second-order SPT predictions, especially for small overdensities. Motivated by this: 
\begin{enumerate}
    \item We have calculated the effect of the density distribution on a general overdensity response function.    \item We have shown that this causes a bias for $\delta_w=0$, even in small volumes, due to the lognormal distribution of densities.    \item We have introduced a phenomenological power-law response function for the position-dependent power spectrum by combining the first-order SPT result with the constrains of the lognormal density distribution, and we have demonstrated its accuracy on cosmological $N$-body simulations.
\end{enumerate}

The new response function is straightforward to calculate using standard tools from cosmological parameters, and it provides an extremely accurate prediction for the local power spectrum. In particular, it is especially useful for low density regions, which includes most of the Universe due to log-normality, where low-order SPT fails catastrophically.

A useful application of our formula would be to correct the measured small-scale power spectrum in zoom-in simulations, with or without the combination of the phase inversion method described by \cite{2016MNRAS.462L...1A}, to reduce the effects of the cosmic variance. Our formulas correct the local power spectrum even for an average density simulation, where the SPT approach predicts no bias.

Our fit predicts a super survey bias of $B(\sigma) \simeq 1 - 2 \sigma^2$ at $z=0$, the accuracy of which was verified in the simulations. This is a consequence of the skewed (lognormal) distribution of the overdensities: a typical sub-volume will be underdense, and the average density sub-volume will have a power spectrum bias of order $-2\sigma^2$. In the Gaussian approximation, the $k_{max}$ of the sub-volume would determine the errors on the overall power spectrum amplitude, and thus this bias would always be significant. In reality, there is a plateau in the power spectrum information due to non-Gaussianity \citep{2006MNRAS.371.1205R, 2006MNRAS.370L..66N} that limits the accuracy that is achievable with two-point statistics. \cite{2014MNRAS.444..994C} identified the plateau $\sigma_{min}$, the minimum achievable super survey variance, as a quadrature sum of the super survey variance, $\sigma_{SS} \simeq \frac{26}{21}\sigma(V)$ (local), and the intra-survey variance, $\sigma_{IS} \simeq P(k_{max})/V$. The former alone is typically larger than the super survey bias identified here, which depends on the square of the variance. Therefore, in most practical cases with large enough sub-volumes of $L \gtrsim 200\textnormal{Mpc}h^{-1}$, the bias will be at the sub-percent level and smaller than the super survey variance.

The main consequence of the lognormal distribution and the power-law response is that the mean density volumes have power spectra that are smaller than the average, full-volume power spectrum.

\section*{Acknowledgements}

This work was supported by the  Ministry of Innovation and Technology NRDI Office grants OTKA NN 129148 and the MILAB Artificial Intelligence National Laboratory Program. IS acknowledges support from the National Science Foundation (NSF) award 1616974.
GR’s research was supported by an appointment to the NASA Postdoctoral Program administered by Oak Ridge Associated Universities under contract with NASA. GR was supported by JPL, which is run under contract by California Institute of Technology for NASA.
We thank A. S. Szalay for insightful suggestions and comments.

%
%

\bibliographystyle{aa}
\bibliography{EmpiricalPowerSpectrumResponse}


\appendix

\section{Approximate $B(\sigma)$ functions}
\label{sec:AppA}

\begin{figure}
    \centering
        \includegraphics[width=0.50\textwidth]{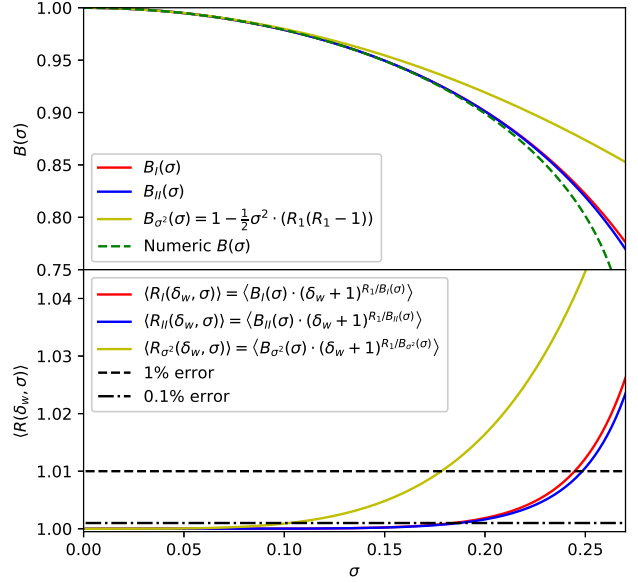}
        \caption{Approximate super survey biases. \textbf{Top:} Approximate  $B_{I}(\sigma)$ and $B_{II}(\sigma)$  solutions with the numeric $B(\sigma)$ function. \textbf{Bottom:} Error in the average of the $R_I(\delta_w,\sigma)$ and $R_{II}(\delta_w,\sigma)$ responses generated from the $B_{I}(\sigma)$ and $B_{II}(\sigma)$ functions, respectively. We have adopted $R_1=47/21+1/3$ for this plot.}
        \label{fig:ApproximateAsigma}
\end{figure}

For a general overdensity response function, $R(k,\delta_w,\sigma)$, the average power spectrum in lognormal density distribution can be calculated as
\begin{equation}
    \left\langle P_w(k,\sigma) \right\rangle= \int\limits_{-1}^{\infty}  R(k,\delta_w,\sigma)P(k) f_{ln}(\delta_w,\sigma)d\delta_w.
    \label{eq:AveragePower}
\end{equation}
Since the average sub-volume spectrum is equal to the full-volume spectrum, Eq.~\ref{eq:AveragePower} can be simplified as
\begin{equation}
    1 = \int\limits_{-1}^{\infty}  R(k,\delta_w,\sigma)f_{ln}(\delta_w,\sigma)d\delta_w.
\end{equation}
With our power-law assumption,
\begin{equation}
    R(\delta_w,\sigma) = B(\sigma)\left(\delta_w+1\right)^{R_1/B(\sigma)},
\end{equation}
for the response function, the integral on the right side of Eq.~\ref{eq:AveragePower} can be solved analytically. The equation then becomes
\begin{equation}
    1 = B(\sigma)\cdot e^{\frac{\sigma^2}{2}\left[\left(\frac{R_1}{B(\sigma)}\right)^2-\frac{R_1}{B(\sigma)}\right]}
    \label{eq:AveragePowerResponse}
.\end{equation}
This transcendental equation has no known analytical solution for $B(\sigma)$, but approximate solutions can be constructed using Taylor series. If we define the
\begin{equation}
    G(B) := B(\sigma)\cdot e^{\frac{\sigma^2}{2}\left[\left(\frac{R_1}{B(\sigma)}\right)^2-\frac{R_1}{B(\sigma)}\right]}-1
\end{equation}
function, Eq~\ref{eq:AveragePowerResponse} can be written as
\begin{equation}
    0=\sum\limits_{n=0}^{\infty} \left.\frac{1}{n!}\frac{d^n G(B)}{dB^n}\right|_{B=1}\cdot\left(B-1\right)^n.
\end{equation}
An approximate solution that uses the first few terms of this Taylor expansion around $B=1$ is expected to be close to the real solution for small $\sigma$ values since $B(\sigma)=1$ when $\sigma=0$. The first two derivatives of $G(A)$ are
\begin{equation}
    \frac{d}{dB} G(B) = \left(\frac{1}{2} {{\sigma }^{2}} \, \left( \frac{R_1}{{{B}}}-\frac{2 {{R_1}^{2}}}{{{B}^{2}}}\right) \, + 1\right){{ e}^{\frac{1}{2} {{\sigma }^{2}}\, \left( \frac{{{R_1}^{2}}}{{{B}^{2}}}-\frac{R_1}{B}\right) }}
\end{equation}

\begin{equation}
    \frac{d^2}{dB^2} G(B) = \left(\frac{1}{4} {{\sigma }^{4}} B\, {{\left( \frac{R_1}{{{B}^{2}}}-\frac{2 {{R_1}^{2}}}{{{B}^{3}}}\right) }^{2}}\, + {{\sigma }^{2}}\,  \frac{ {{R_1}^{2}}}{{{B}^{3}}} \, \right){{ e}^{\frac{1}{2} {{\sigma }^{2}}\, \left( \frac{{{R_1}^{2}}}{{{B}^{2}}}-\frac{R_1}{B}\right) }}.
\end{equation}
Using these, different orders of polynomial equations can be constructed for $B(\sigma)$. For $n\leq1$, the first-order equation can be written as
\begin{equation}
\begin{split}
    &e^{\frac{1}{2}\sigma^2\left(R_1^2-R_1\right)}-1 + \\
    &+\left[\left(\frac{1}{2}\sigma^2\left(R_1-2R_1^2\right) + 1\right)e^{\frac{1}{2}\sigma^2\left(R_1^2-R\right)}\right]\left(B_{I}(\sigma)-1\right) = 0,
    \label{eq:FirstOrderAeq}
\end{split}
\end{equation}
where the first-order solution is
\begin{equation}
    B_{I}(\sigma) = 1 - \frac{1-e^{-\frac{1}{2}\sigma^2\left(R_1^2-R_1\right)}}{\left(\frac{1}{2}\sigma^2\left(R_1-2R_1^2\right) + 1\right) }.
    \label{eq:FirsOrderA}
\end{equation}
The $n\leq2$ equation is
\begin{equation}
    \begin{split}
        &e^{\frac{1}{2}\sigma^2\left(R_1^2-R_1\right)}-1 +\\
        &+ \left[\left(\frac{1}{2}\sigma^2\left(R_1-2R_1^2\right) + 1\right)e^{\frac{1}{2}\sigma^2\left(R_1^2-R_1\right)}\right]\left(B_{II}(\sigma)-1\right) +\\
        &+ \frac{1}{2}\left[\left( \frac{1}{4}\sigma^4\left(R_1-2R_1^2\right)^2 + \sigma^2 R_1^2 \right)e^{\frac{1}{2}\sigma^2\left(R_1^2-R_1\right)}\right](B_{II}(\sigma) - 1)^2 = 0,
    \end{split}
\end{equation}
and the solution is
\begin{equation}
    \begin{split}
        &B_{II}(\sigma) = 1 -  \frac{1}{{\left[\left( \frac{1}{4}\sigma^4\left(R_1-2R_1^2\right)^2 + \sigma^2 R_1^2 \right)e^{\frac{1}{2}\sigma^2\left(R_1^2-R_1\right)}\right]}} \cdot \\
        &\cdot\left(\left[\left(\frac{1}{2}\sigma^2\left(R_1-2R_1^2\right) + 1\right)e^{\frac{1}{2}\sigma^2\left(R_1^2-R_1\right)}\right] + \right.\\
        &+ \left(\left[\left(\frac{1}{2}\sigma^2\left(R_1-2R_1^2\right) + 1\right)e^{\frac{1}{2}\sigma^2\left(R_1^2-R_1\right)}\right]^2-\right.\\
        &\left.\left.-2\left[\left( \frac{1}{4}\sigma^4\left(R_1-2R_1^2\right)^2 + \sigma^2 R_1^2 \right)e^{\frac{1}{2}\sigma^2\left(R_1^2-R_1\right)}\right]\left(e^{\frac{1}{2}\sigma^2\left(R_1^2-R_1\right)}-1\right)\right)^{\frac{1}{2}} \right).
    \end{split}
\end{equation}
The first- and second-order solutions can be seen in the top panel of Fig.~\ref{fig:ApproximateAsigma}. The precision of these solutions can be checked by substituting $B(\sigma)$ back into Eq.~\ref{eq:AveragePowerResponse}. This is plotted in the bottom panel of Fig.~\ref{fig:ApproximateAsigma}.

\end{document}